\documentclass[letterpaper,twocolumn,10pt]{article}
\usepackage{usenix-2020-09}

\usepackage{tikz}
\usepackage{amsmath}
\usepackage{comment}
\usepackage{titling}
\usepackage{amsfonts}
\usepackage[normalem]{ulem}
\usepackage[htt]{hyphenat}
\usepackage{braket}
\usepackage[ruled,vlined]{algorithm2e}
\usepackage{listings}
\usepackage{subcaption}

\usepackage[inline]{enumitem}
\setlist{noitemsep,topsep=0pt,parsep=0pt,partopsep=0pt}

\usepackage[subtle]{savetrees}

\usepackage{setspace}
\usepackage[margin=5pt,font={stretch=0.9}]{caption}

\usepackage{libertine}
\usepackage{grumble}

\newcommand{\projecttitle}{QOS}
\newcommand{\myparagraph}[1]{\noindent{\bf {#1}.}}
\newcommand*\circled[1]{\tikz[baseline=(char.base)]{
            \node[shape=circle,draw,inner sep=1pt] (char) {#1};}}

\newcommand{\setFlag}{false} 
\newcommand{\coloredText}[1]{%
  \ifthenelse{\equal{\setFlag}{true}}%
  {\textcolor{blue}{#1}}%
  {#1}%
}

\begin{document}

\date{}




\title{\Large \bf \projecttitle: A Quantum Operating System}



\author{
Emmanouil Giortamis \quad Francisco Romão \quad Nathaniel Tornow \quad Pramod Bhatotia \\
Technical University of Munich
}

\maketitle

\begin{abstract}

Quantum computers face challenges due to hardware constraints, noise errors, and heterogeneity, and face fundamental design tradeoffs between key performance metrics such as \textit{quantum fidelity} and system utilization. 
This substantially complicates managing quantum resources to scale the size and number of quantum algorithms that can be executed reliably in a given time. 

We introduce \projecttitle{}, a cloud operating system for managing quantum resources while mitigating their inherent limitations and balancing the design tradeoffs of quantum computing. \projecttitle{} exposes a hardware-agnostic API for transparent quantum job execution, mitigates hardware errors, and systematically multi-programs and schedules the jobs across space and time to achieve high quantum fidelity in a resource-efficient manner. To achieve this, it leverages two key insights: First, to maximize utilization and minimize fidelity loss, some jobs are more compatible than others for multi-programming on the same quantum computer. Second, sacrificing minimal fidelity can significantly reduce job waiting times.


We evaluate \projecttitle{} on real quantum devices hosted by IBM, using 7000 real quantum runs of more than 70.000 benchmark instances. We show that the \projecttitle{} achieves 2.6--456.5$\times$ higher fidelity, increases resource utilization by up to 9.6$\times$, and reduces waiting times by up to 5$\times$ while sacrificing only 1--3\% fidelity, on average, compared to the baselines.



\end{abstract}

\section{Introduction}

Quantum computing promises to solve computationally intractable problems with classical computers~\cite{arute2019quantum, daley2022practical, grover1996fast, shor1999polynomial}. Thanks to remarkable technological advances in materials science and engineering, quantum hardware has become a reality in the form of quantum processing units (QPUs) \cite{siddiqi2021engineering, hanneke2010realization, dicarlo2009demonstration}. Interestingly, QPUs are now readily available in a quantum-as-a-service fashion offered by all major cloud providers~\cite{ibmQuantum, aws-quantum, google-quantum, azure-quantum}.

However, QPUs present fundamentally unique hardware-level challenges that cannot be directly mapped to classical accelerator-oriented computing (we empirically detail these hardware challenges in \S~\ref{sec:design-challenges}). In particular, QPUs operate in the NISQ-fashion (Noisy Intermediate-Scale Quantum \cite{preskill2018quantum}), leading to strict \textit{hardware constraints} and a {\em non-deterministic} computing platform \cite{ravi2022quantum, patel2020experimental}.

More specifically, QPUs are inherently noisy and small in computational capacity, which limits the size of the problems they can solve \cite{preskill2018quantum}. Second, the degree of noise differs across QPUs and time, even of identical architecture and model, making it difficult to decide which QPUs should execute a quantum program with good performance \cite{ravi2022quantum}. 
In addition, we can not trivially multi-program multiple quantum jobs on the same QPU to increase utilization without them interfering with each other in undesirable and unpredictable ways \cite{murali2020software}, severely degrading performance \cite{liu2021qucloud}. Last, scheduling quantum jobs on heterogeneous noisy QPUs exhibits a fundamental tension between quantum performance metrics, such as \textit{fidelity} \cite{fidelity-qiskit}, and classical performance metrics, such as utilization and load balance (i.e., resource efficiency).

Unfortunately, these problems are amplified by the current state of software, where QPUs are managed through rudimentary interfaces, despite the ever-increasing demand for these scarce resources \cite{ravi2022quantum, patel2020experimental, qiskit-trillion-circuits}. \coloredText{Researchers have proposed specialized approaches to address some of the aforementioned OS and QPU challenges individually, for instance, fidelity without runtime support \cite{Ayanzadeh2023frozenQbits},  basic multi-programming via FIFO or random job selection \cite{das2019a}, or heuristics-based scheduling \cite{ravi2021adaptive}.
Unfortunately, these systems are designed in isolation, are tightly coupled to specific policies, and lack the architectural flexibility and interoperability needed for cross-stack coordination.}

\coloredText{At the same time, in industry platforms like IBM Cloud ~\cite{ibmQuantum} and AWS Braket \cite{aws-quantum}, multi-programming is not even supported, and users must improve fidelity and select QPUs \textit{manually}. However, the noisy, scarce, and heterogeneous nature of QPUs naturally motivates users to select the best-performing QPUs, further amplifying the already high QPU load imbalance and queue waiting times \cite{ravi2022quantum, patel2020experimental}.}


\coloredText{In total, there is still no cloud operating system that addresses quantum job and resource management in a unified and systematic way, i.e., an architecture that supports composable, cross-layer mechanisms designed specifically for the constraints of quantum computing. Such a system must integrate error mitigation techniques to improve fidelity, fidelity-aware performance estimation to handle the spatiotemporal variability of QPUs, multi-programming to increase utilization through compatibility-aware co-location of quantum jobs, and a scheduler that balances QPU load and reduces queue times—all under a single, unified stack.}

To fill this gap, we propose \projecttitle{}, a cloud operating system for holistically tackling quantum computing challenges \coloredText{through a modular, cross-layer architecture. At the core of \projecttitle{} lies the \textit{Qernel} abstraction—the common denominator that enables the seamless composition of diverse system mechanisms and facilitates cross-layer optimizations.}

\projecttitle{} abstracts away the underlying complexity of quantum resource management and systematically explores the associated tradeoffs of quantum computing. To achieve this, \projecttitle{} exposes hardware-agnostic APIs and comprises four main components: (1) the \textit{error mitigator}, a component \coloredText{that composes complementary techniques to increase fidelity by mitigating hardware noise}, (2) the \textit{estimator}, \coloredText{which predicts fidelity performance across heterogeneous QPUs to guide scheduling decisions}, (3) the \textit{multi-programmer} that \coloredText{bundles compatible jobs on the same QPU to improve QPU utilization while minimizing fidelity loss}, and (4), the \textit{scheduler}, \coloredText{which performs fidelity-aware, multi-objective job scheduling to reduce queueing latency and balance QPU load}.

\begin{figure}
    \centering
    \includegraphics[width=\columnwidth]{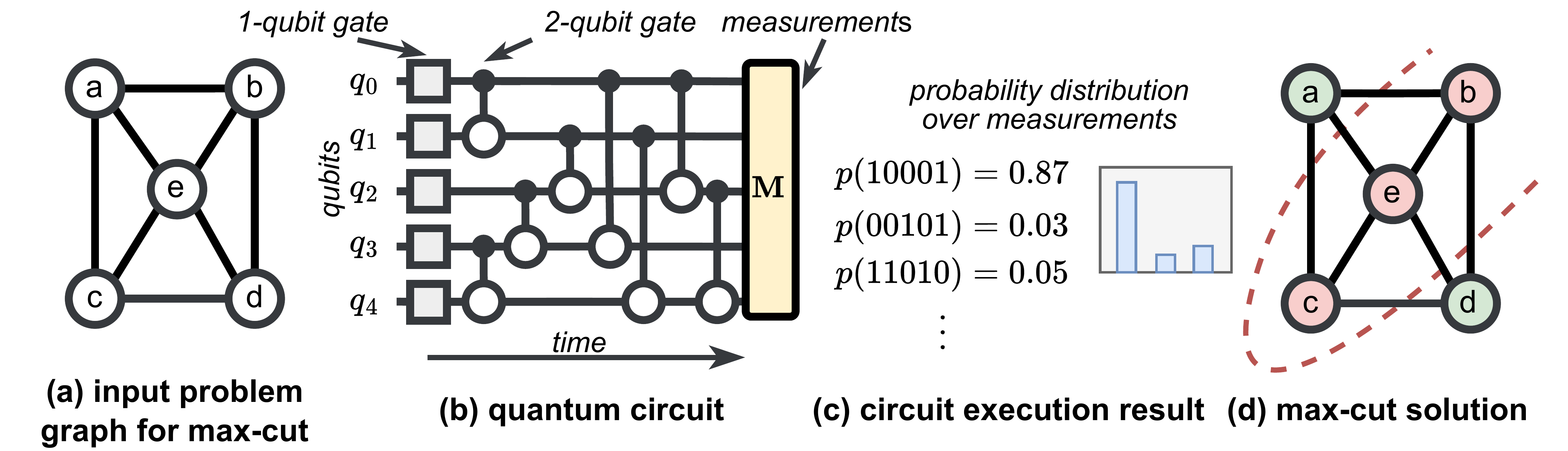}
    \caption{Foundational example (\S~\ref{sec:background:101}) {\em {\bf (a)} Input graph to max-cut. {\bf (b)} A quantum circuit encoding the max-cut formulation for the graph. {\bf (c)} The execution result is a probability distribution of bitstrings. {\bf (d)} The result is interpreted as a max-cut between vertices $\{a, d\}$ and $\{b,c,e\}$. }}
    \label{fig:qaoa-background}
\end{figure}

We implement \projecttitle{} in Python by building on the Qiskit framework \cite{Qiskit}. 
We evaluate \projecttitle{} on IBM's 27-qubit QPUs \cite{ibmQuantum}, using a dataset of more than 7000 quantum runs and 70.000 state-of-the-art quantum benchmark instances used in popular quantum algorithms \cite{tomesh2022supermarq, quetschlich2022mqt, li2021qasmbench}. Our evaluation shows that the error mitigator improves execution fidelity by 2.6--456.5$\times$ on average, depending on the problem size (\S~\ref{sec:evaluation:compiler}), the estimator correctly identifies high-fidelity QPUs (\S~\ref{evaluation:results:estimator}), the multi-programmer fidelity by 1.15--9.6$\times$ for a target utilization (\S~\ref{evaluation:results:multi-programmer}), and the scheduler reduces the waiting times by 5$\times$ while sacrificing at most 3\% of fidelity (\S~\ref{evaluation:results:scheduler}).

\myparagraph{Contributions} Our main contributions include:
\begin{enumerate} 
    \item \coloredText{ \projecttitle{} is the first quantum operating system to holistically address the challenges of quantum computing. Its modular architecture and Qernel abstraction enforce a clean separation of mechanism and policy, enabling flexible integration of evolving techniques without modifying system interfaces.}

    \item \coloredText{\projecttitle{} enables both cross-layer and intra-layer optimizations through its end-to-end system design. \projecttitle{}'s components synergistically optimize fidelity and resource efficiency while supporting composable techniques within individual layers.    
    }

    \item \coloredText{\projecttitle{} systematically explores key tradeoffs in quantum computing: Fidelity versus overheads in error mitigation and performance estimation, as well as fidelity versus QPU utilization and waiting times in multi-programming and scheduling, respectively.}

    \item \coloredText{\projecttitle{} introduces new ideas across the quantum software stack. These include the \textit{Qernel} abstraction as a unifying execution unit, the composable error mitigation pipeline combining complementary techniques, a multi-programming model that quantifies the \textit{compatibility} and \textit{effective} utilization of co-located jobs, and the first fidelity-aware multi-objective scheduler.    
    }

\end{enumerate}

\begin{figure} [t]
    \centering
    \includegraphics[width=\columnwidth]{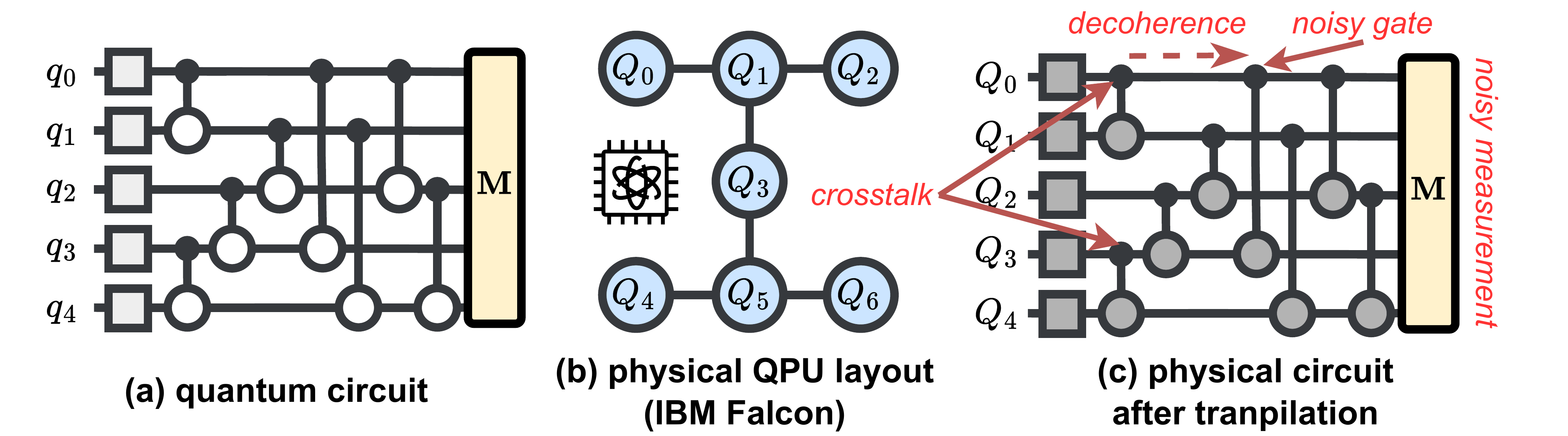}
    \caption{Technical Foundations (\S~\ref{sec:background:foundations}) {\em {\bf (a)} The quantum circuit of Figure \ref{fig:qaoa-background}. {\bf (b)} The physical layout of an IBM Falcon QPU. {\bf (c)} The transpiled circuit with the QPU's noise sources. }}
    \label{fig:background-detail}
\end{figure}

\section{Background}
\label{sec:background}

\subsection{Quantum Computing 101}
\label{sec:background:101}

Quantum computers solve \textit{specific} computationally hard problems exponentially faster than classical computers by leveraging the quantum mechanics principles of \textit{superposition} and \textit{entanglement}. Specifically, the basic units of quantum computation are \textit{qubits}, which during quantum computation are both 0 and 1 at the same time (recall Schrodinger's cat experiment \cite{schrondigers-cat}), and when entangled, they can interact with each other even over large distances \cite{ma2012quantum}.

To solve NP-hard problems (e.g., maximum-cut) with quantum computers, we use algorithms such as the Quantum Approximate Optimization Algorithm (QAOA) \cite{farhi2014quantum}. This algorithm is considered practical for today's quantum hardware capabilities and influential towards achieving quantum advantage \cite{riste2017demonstration}. 

\myparagraph{Foundational example}
Figure \ref{fig:qaoa-background} shows how QAOA solves an example max-cut problem for an input graph \textbf{(a)}. First, the graph is encoded as a \textit{quantum circuit}. Quantum circuits comprise qubits and quantum gates, akin to logical gates in classical circuits (e.g. NOT, XOR), which are applied over time (from left to right). In the max-cut case, each graph vertex corresponds to a qubit and each edge to a 2-qubit gate in the circuit \textbf{(b)}. At the end of the circuit, we measure each qubit to read its value, which gives bitstrings as output. Notably, measurements collapse the superposition state to a definite binary value.

\begin{figure*}[t] 
    \centering
    \includegraphics[width=\textwidth]{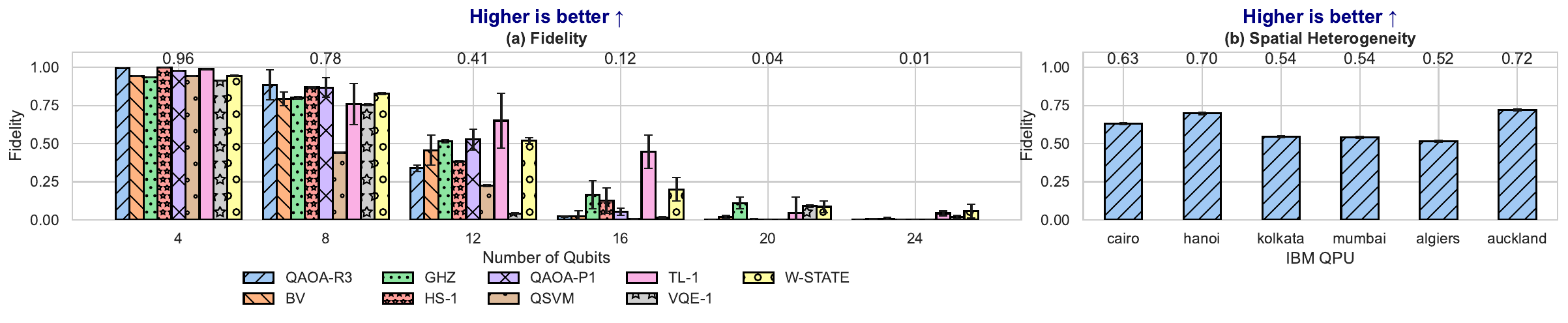}
    \caption{
    \textbf{(a)} Challenge \#1, Fidelity (\S~\ref{sec:challenges:scalability}). 
        {\em Impact of the number of qubits (circuit size) on fidelity. There is an average 98.9\% reduction in fidelity from 4 to 24 qubits.
        }
    \textbf{(b)} Challenge \#2, Spatial heterogeneity (\S~\ref{sec:challenges:spatial_and_temporal}).
        {\em Fidelity of a 12-qubit GHZ circuit on different IBM QPUs. There is a 38\% fidelity difference from best to worst QPU.
        }    
    }
    \label{fig:scalability_spatial}
\end{figure*}

Since quantum mechanics is inherently probabilistic, the bitstring we get is truly random, but due to the algorithm structure, it follows a \textit{probability distribution}, which ultimately renders quantum computers useful. Thus, we execute the circuit in many trials (\textit{``shots''}), with each trial providing a specific bitstring from the qubit measurements.
The solution of the quantum calculation is, therefore, a probability distribution over all possible bitstrings of the measured qubits, where high probability maps to the solution, while low ($\sim0$) does not represent a solution \textbf{(c)}. In our example, the bitstring \texttt{10001} represents the max-cut solution that contains the partitions \{a, d\} and \{b, c, e\} \textbf{(d)}.

\subsection{Technical Foundations}
\label{sec:background:foundations}

\myparagraph{Execution Model}
The technology and engineering required to build QPUs renders them an expensive resource, i.e., there are less than 100 QPUs globally offered in the cloud in a quantum-as-a-service fashion \cite{ibmQuantum, azure-quantum, google-quantum, aws-quantum}. To run quantum programs, users typically write circuit-level code (Figure~\ref{fig:background-detail} (a)), which then \textit{transpile} on the QPU to make it executable, send it to the cloud for execution, and finally get the results back. Specifically, the transpilation process performs three key steps: (1) converting the gates of the circuit to the native gate set of the QPU, (2) mapping the logical qubits of the circuit to the physical qubits of the QPU, (3) routing the qubits to the physical qubits with restrictive connectivity by inserting additional costly gates. Figure \ref{fig:background-detail} (b) shows the physical layout of an IBM Falcon QPU. Vertices are the physical qubits, and the edges capture their connectivity, i.e., between which qubits we can apply 2-qubit gates. Figure \ref{fig:background-detail} (c) shows the physical circuit after transpilation with the QPU's noise characteristics, which we detail next.

\myparagraph{QPU characteristics}
Today's QPUs are described as noisy intermediate-scale quantum (NISQ) devices \cite{preskill2018quantum} since they exhibit low qubit numbers (e.g., up to a few 100s \cite{ibmQuantum}) and are susceptible to hardware and environmental noise. Specifically, when measuring a qubit, there is a chance to read the opposite value, and when applying gates, there is a chance the gate performs a wrong operation \cite{google-nisq-properties}. On top of that, when qubits are left idle (no gates applied) for more than a few hundred microseconds, the superposition \textit{decoheres} to the $\ket{0}$ state \cite{klimov2018fluctiations}, similar to resetting a register to $0$. Lastly, qubits destructively interfere with each other via \textit{crosstalk} effects \cite{murali2020software}. Figure~\ref{fig:background-detail} (c) shows qubits $Q_0$ and $Q_3$ that influence each other via crosstalk, noisy gates, qubit $Q_0$ that is left idle for long enough to decohere, and noisy measurements.

\myparagraph{QPU heterogeneity}
Additionally, QPUs are vastly \textit{heterogeneous} across space and time, unlike classical accelerators. Across space, QPUs vary in terms of technology, e.g., superconducting qubits \cite{siddiqi2021engineering, dicarlo2009demonstration} or trapped ions \cite{haffner2008quantum}, architectures of the same technology, e.g., Falcon or Osprey superconducting QPUs \cite{ibmQuantum}, and noise properties (formally called \texttt{noise model}) even for the same architecture \cite{google-nisq-properties}, e.g., two identical QPUs exhibit different noise errors, etc. Across time, the QPUs are \textit{calibrated} regularly to maintain their performance \cite{wittler2021integrated, tornow2022minimum, ibmq-calibration}, a process that generates \textit{calibration data}. These data quantify the noise errors and change unpredictably after each calibration cycle.

\myparagraph{Performance metric}
To measure the quality of a circuit execution on NISQ QPUs, we use the \textbf{fidelity} metric \cite{fidelity-qiskit}, which measures the similarity between the noisy probability distribution and the ideal probability distribution that noiseless, ideal QPUs can obtain.
\coloredText{Given two probability distributions over measurement outcomes, the ideal distribution $P_{ideal}$ and the noisy distribution $P_{noisy}$, fidelity is computed as: $F(P_{ideal}, P_{noisy}) = (\sum_{i}^{} \sqrt{P_{ideal}(i) * P_{noisy}(i)})^2$.} Fidelity is a number in the $[0, 1]$ range, where a higher fidelity means a better quality result.

\myparagraph{Circuit properties}
\coloredText{Execution fidelity is impacted by the aforementioned QPU noise model and the \textit{circuit properties} that describe the circuit's complexity and, thus, susceptibility to errors. Circuit \textit{width} refers to the number of logical qubits involved in the circuit and typically also the physical (H/W) qubits required to execute it. Circuit \textit{depth} denotes the largest number of layers of gates applied to the qubits and quantifies the circuit runtime. A larger depth typically indicates a higher chance for decoherence errors. Last, the number of \textit{non-local gates} refers to the gates that act on non-adjacent qubits, which are particularly susceptible to errors in NISQ devices. \textit{Generally, larger circuits (larger width, depth, or number of non-local gates) equals lower fidelity}.
}

\begin{figure*}[ht] 
    \centering
    \includegraphics[width=\textwidth]{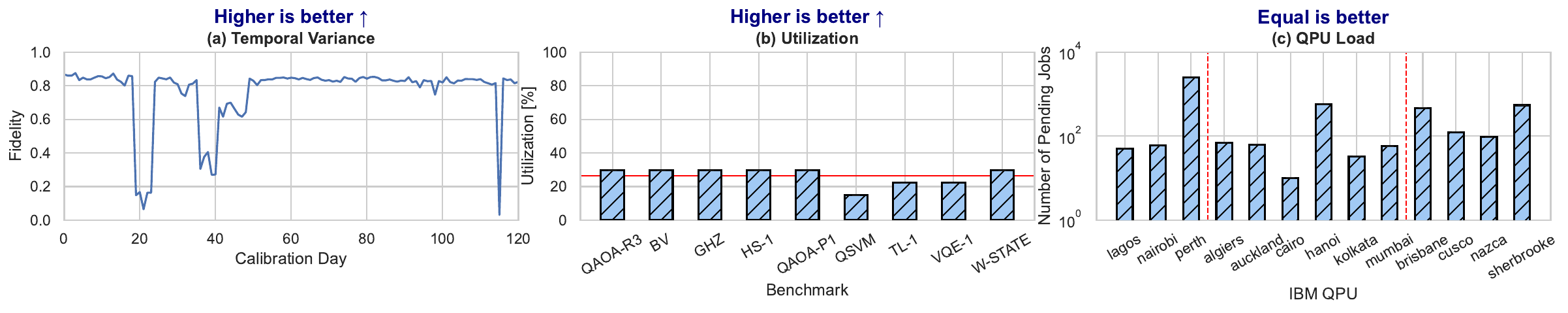}
    \caption{
    \textbf{(a)} Challenge \#2, Temporal variance (\S~\ref{sec:challenges:spatial_and_temporal}).
         {\em Fidelity of a 6-qubit GHZ circuit on IBM Perth, across 120 calibration days. There are 20 pairs of days with more than 5\% difference in fidelity.
        }
    \textbf{(b)} Challenge \#3, Utilization (\S~\ref{sec:challenges:utilization}).
        {\em Maximum utilization achieved on a 27-qubit QPU for nine benchmarks while maintaining at least 0.75 fidelity. The average utilization is $26.3\%$, and the max is $29.6\%$.
        }
    \textbf{(c)} Challenge \#4, QPU Load (\S~\ref{sec:challenges:qpu_load}).
         {\em Number of pending jobs on different IBM QPUs. The groups separated by vertical red lines indicate QPUs of the same size. There is up to 57$\times$ difference in number of jobs between QPUs of the same size.
        }
    }
    \label{fig:temporal_util_loads}
\end{figure*}

\myparagraph{Circuit feature vectors}
\coloredText{ While circuit properties are useful, sometimes they fail to characterize the structural and computational properties of quantum circuits. Thus, we can use the Supermarq feature vectors, a set of six metrics proposed by Tomesh et al. \cite{tomesh2022supermarq} with the initial goal of quantifying quantum benchmark coverage. These features quantify the parallelism and speedup potential, the lower bound on execution time, QPU connectivity requirements, entanglement (i.e. complexity), and susceptibility to decoherence and measurement errors. 
}

\section{Motivation and Key Ideas}
\label{sec:design-challenges}

To motivate \projecttitle{}, we present a set of \textit{unique} challenges that distinguish QPUs from classical accelerators. 
We categorize our findings into four challenges that must be addressed to improve the practicality of quantum computing: fidelity, utilization, spatial and temporal heterogeneities, and load imbalance.
The experimental methodology used is the same for the final system evaluation and is explained in detail in \S~\ref{sec:evaluation:experimental_methodology}.

\subsection{Fidelity}
\label{sec:challenges:scalability}

Executing quantum programs with high fidelity is challenging since QPUs are characterized by relatively small numbers of qubits and noise, which leads to computation errors (\S~\ref{sec:background:foundations}). As the number of qubits and gates in a quantum circuit increases, the noise errors accumulate and the overall fidelity decreases.

\myparagraph{Results} Our results are highlighted in Figure \ref{fig:scalability_spatial} (a). The \textit{x} axis shows the circuit size as the number of qubits while the \textit{y} axis shows the fidelity, where higher is better. The experiment is run on the IBM Kolkata 27-qubit QPU. For each increase in qubits, the average fidelity decreases, up to $98.9\%$ from 4 to 24 qubits. Moreover, it is physically impossible to run circuits with a size larger than 27 qubits, since we cannot map them.

\myparagraph{Implication} NISQ devices are limited due to size and noise and, therefore, cannot be practically used for large quantum circuits, either logically because the circuit doesn't fit in the device or the execution results would be degraded from noise-induced errors, which translates to low fidelity.

\subsection{Spatial and Temporal Heterogeneity}
\label{sec:challenges:spatial_and_temporal}

In the classical domain, two identical CPUs perform similarly for all applications at each point in time. In contrast, QPUs exhibit differences in the layout and connectivity of qubits \cite{gyongyosi2019a} and variations in noise errors even for QPUs of the same model, which leads to spatial performance variance. 
Moreover, QPUs are calibrated regularly (\S~\ref{sec:background:foundations}), and after each calibration, the noise properties change \cite{ravi2022quantum}. As a result, the execution fidelity can vary across different calibration cycles, leading to temporal performance variance.

\myparagraph{Results}
Figure \ref{fig:scalability_spatial} (b) shows a 12-qubit GHZ circuit's fidelity on different IBM QPUs. Fidelity varies across the QPUs, with a maximum difference of 38\% from best to worst. Notably, all six QPUs are of the same model (Falcon r5.11).

Figure \ref{fig:temporal_util_loads} (a) shows a 6-qubit GHZ circuit's fidelity over 120 calibration days executed on the IBM Perth 7-qubit QPU, where each data point represents a single day's fidelity. The largest single-day difference in fidelity is 96.5\%, and there are 20 instances of a single-day fidelity drop of more than 5\%. Note that there is no way to predict a QPU's future calibration data to expect such performance differences. 

\myparagraph{Implications}
Due to structural differences across QPUs, quantum circuits perform differently across them. 
Additionally, there is a high degree of temporal performance variance across calibration cycles, as the fidelity might change significantly from day to day with no discernible pattern. 

\subsection{Utilization}
\label{sec:challenges:utilization}

The fidelity of circuits decreases as their size increases (\S~\ref{sec:challenges:scalability}), and as a result, it becomes more challenging to utilize a QPU effectively. In contrast to the classical domain, where a CPU can be fully utilized, to get high-fidelity results in the quantum domain, we necessarily under-utilize QPUs.

\myparagraph{Results} Figure \ref{fig:temporal_util_loads} (b) shows the maximum utilization of the IBM Kolkata 27-qubit QPU for nine benchmarks while targeting at least 0.75 fidelity. No benchmark exceeds 30\% utilization, while the average is 26.3\%. Higher fidelity targets would yield even lower utilization and vice-versa.

\myparagraph{Implications}
There is a tradeoff between QPU utilization and fidelity. In general, the lower utilization, the higher fidelity, and vice-versa. In contrast to the classical domain, the tension between these metrics is vastly larger.

\subsection{QPU Load Imbalance}
\label{sec:challenges:qpu_load}

The quantum cloud faces QPU load imbalance. The root cause is spatiotemporal heterogeneity (\S~\ref{sec:challenges:spatial_and_temporal}), combined with the manual QPU selection offered by the current quantum cloud model \cite{ibmQuantum}. These naturally motivate users to select the highest fidelity QPU(s) for that calibration cycle \cite{ravi2022quantum}.

\myparagraph{Results}
Figure \ref{fig:temporal_util_loads} (c) shows the average number of pending jobs for different IBM QPUs across October 2023. The groups of QPUs (separated by the red dashed line) have a size of 7, 27, and 127 qubits, respectively. There is a 49$\times$, 57$\times$, and 5.7$\times$ maximum load difference across the groups, respectively.

\myparagraph{Implications}
Load imbalance leads to long waiting times for the users and thus, low quality of service. However, the performance difference does not always justify the load differences between QPUs. For instance, the 12-qubit GHZ circuit in Figure \ref{fig:scalability_spatial} (b) performs 1.1$\times$ better on IBM Hanoi than IBM Cairo, yet the former exhibits 57$\times$ higher load.

\section{Overview} 
\label{sec:overview:overview}

We propose \projecttitle{}, a cloud operating system for quantum computing. In this context, we define a quantum cloud operating system as a systems software layer that transparently manages quantum jobs and resources efficiently w.r.t. job fidelity and waiting times (user's goals) and QPU utilization and load-balance (cloud operator's goals). 
Therefore, \projecttitle{} strives for three design goals: (1) \projecttitle{}'s \coloredText{architecture should be general and modular, cleanly separating mechanism from policy while enabling both cross-layer coordination and within-layer optimizations}. (2) \projecttitle{} should enable the execution of large quantum jobs with high fidelity. (3) \projecttitle{} should be resource efficient by achieving high QPU utilization and balancing QPU load to minimize queue waiting times.

\begin{figure}[t]
    \centering
    \includegraphics[width=\columnwidth]{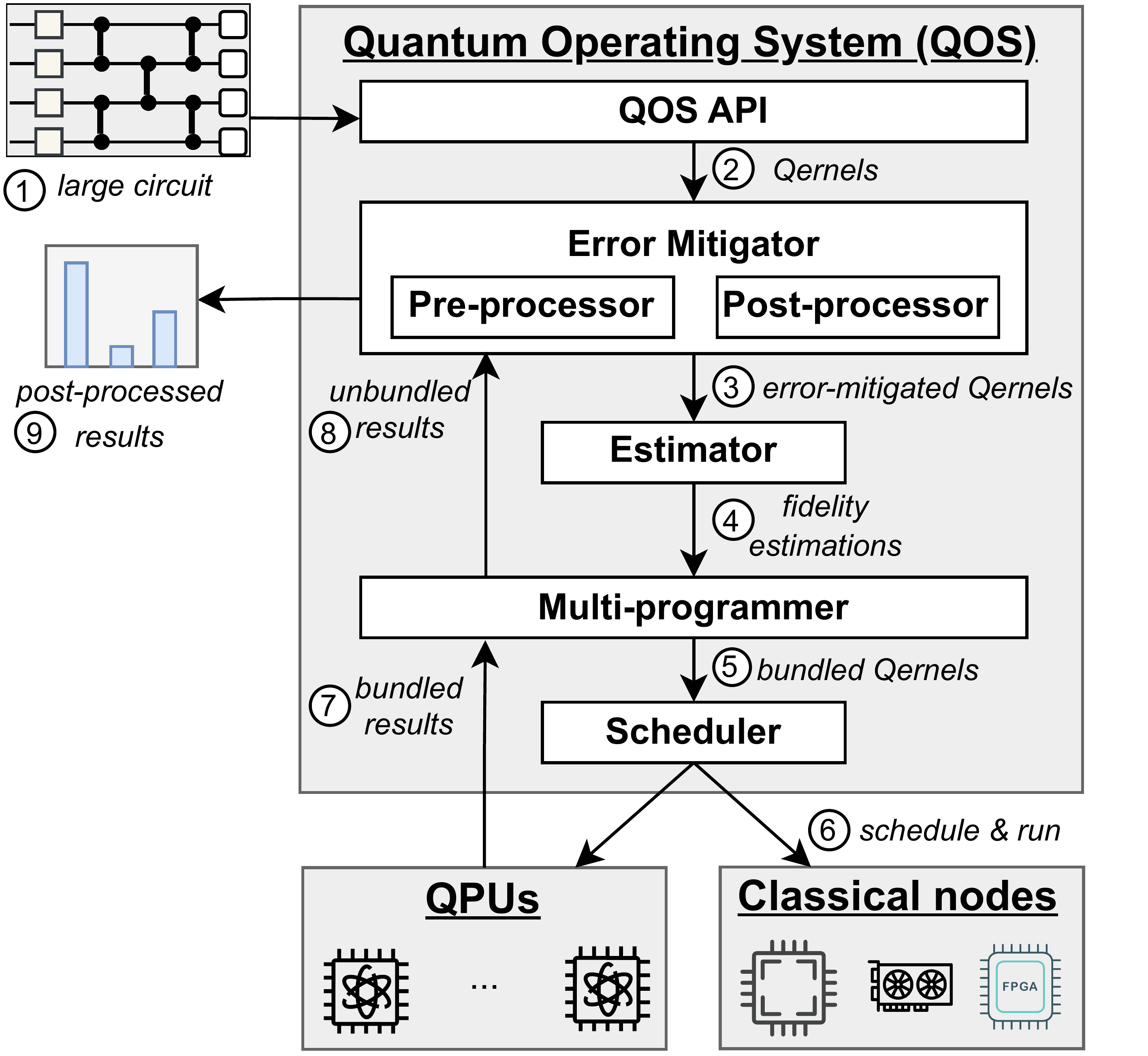}
    \caption{ \projecttitle{} overview (\S~\ref{sec:overview:overview}): {\em
    \projecttitle{} consists of four main components: the error mitigator, estimator, multi-programmer, and scheduler.}}
    \label{fig:overview}
\end{figure}

\subsection{The \projecttitle{} Architecture}
\label{sec:overview:architecture}

Figure \ref{fig:overview} shows the overview of our system's design. \projecttitle{} comprises a layered architecture that consists of an API and four main components: the error mitigator, the estimator, the multi-programmer, and the scheduler, which we detail next. 

\coloredText{\myparagraph{Qernel abstraction}
\projecttitle{} implements a wide range of mechanisms with different abstraction requirements, from error mitigation to scheduling level.
To enable the composability of these mechanisms in a unified architecture, we propose the \textit{Qernel} abstraction that acts as a common denominator for the \projecttitle{} mechanisms to apply their policies.}

\myparagraph{\projecttitle{} API}
The \projecttitle{} API abstracts away the underlying complexity of the noisy and heterogeneous quantum resources by exposing hardware-agnostic functions for configurable quantum job execution. \coloredText{The API transforms the user's input circuits into Qernels for \projecttitle{} to apply its mechanisms.} 

\myparagraph{Error mitigator}
To increase the fidelity of quantum programs, the error mitigator applies pre- and post-processing error mitigation techniques. However, there exists an abundance of such techniques, and they typically incur high runtime overheads that scale with the amount of error mitigation used. Thus, the error mitigator automatically selects and composes a subset of techniques given budget constraints to limit the associated runtime costs.

\myparagraph{Estimator}
To make informed scheduling decisions in the landscape of QPU spatiotemporal performance variance, we require accurate fidelity estimations that do not depend on expensive simulations or trial runs. The estimator leverages analytical models to predict the execution fidelity on the underlying QPUs in a scalable manner.

\myparagraph{Multi-programmer}
There is an inherent tradeoff between QPU utilization (proxy of program size) and fidelity. To increase QPU utilization without significantly compromising fidelity, the multi-programmer spatially multiplexes multiple quantum programs on a single QPU with careful consideration for fidelity performance loss. 

\myparagraph{Scheduler}
There is a fundamental tradeoff between job waiting times and fidelity that arises from the underlying spatial performance variance. The scheduler trades minimal fidelity for significant waiting time improvement by optimizing for this conflicting objective.

\subsection{\projecttitle{} Programming Model}
\label{sec:overview:programming_model}

We introduce the \projecttitle{} programming model designed to abstract away the underlying complexity of managing heterogeneous and noisy quantum resources. \projecttitle{} exposes hardware-agnostic APIs and leverages the Qernel unified abstraction that acts as a common denominator across its components to enable the application of its diverse mechanisms.

\myparagraph{\projecttitle{} APIs}
Table \ref{tab:api} shows the \projecttitle{} APIs that abstract away quantum job execution and resource management. To run a quantum circuit, users simply call \texttt{run} with the circuit and configuration options, such as the error mitigation budget, which controls the runtime overheads and thus cost (in \$). This function returns a unique job id, $jID$. Users can call \texttt{result} and \texttt{status} with the returned $jID$ to retrieve the execution results and status, respectively. Last, users call \texttt{backends} and \texttt{backend\_props} to retrieve the available QPUs and their properties. 

\begin{table}[t]
\fontsize{8}{9}\selectfont 
\caption{\projecttitle{} programming API.} 
\vspace{-9pt}

\begin{center}
\begin{tabular}{ |c|c| }
 \hline
 \bf{\projecttitle{} API} & {\bf Description} \\ \hline
 \texttt{run(circ, cnfgs)} & Run circuit with config. options. \\
 \texttt{results(jID)} & Retrieve the job results.  \\
 \texttt{status(jID)} & Retrieve the job status. \\ 
  \texttt{backends()} & Retrieve the available backends. \\
 \texttt{backend\_props(bID)} & Retrieve the backend properties. \\
  \hline

\end{tabular}
\end{center}
\vspace{-5pt}
\label{tab:api}
\end{table}

\myparagraph{The Qernel abstraction}
\label{sec:qernel}
\label{sec:qernel:properties}
The Qernel implements data structures
that store the static and dynamic properties of the quantum job.
Specifically, to apply error mitigation techniques optimally, we must leverage the characteristics of the quantum circuit, such as the circuit's size (number of qubits), depth, number and types of 2-qubit gates, the number of measurements, etc. Additionally, we include the six Supermarq features vectors \cite{tomesh2022supermarq} since they are potentially useful for heuristic-based optimizations or regression-based prediction models \cite{quetschlich2024mqtpredictor}. This information is described as the static properties of the Qernel. The dynamic properties include the Qernel's execution status (\texttt{done, failed, running, scheduled}), the estimator's output, i.e., fidelity estimations (\S~\ref{sec:qos:estimator}), and the final post-processed results.

\myparagraph{System workflow}
Users submit a large circuit using the \projecttitle{} API \circled{1}. Then, \projecttitle{} transforms the circuit into a Qernel, the common denominator of \projecttitle{}, and submits it to the error mitigator \circled{2}. This component outputs error-mitigated Qernels and submits them to the estimator \circled{3}. The estimator predicts the fidelity of running the Qernel(s) on the QPUs to guide scheduling \circled{4}. Next, the multi-programmer bundles Qernels with low utilization and sends them to the scheduler \circled{6}, who schedules and runs the bundled Qernels, optimizing for maximal fidelity and minimal waiting times \circled{6}. After the execution, the bundled results are unbundled by the multi-programmer into separate results and are sent to the error mitigator for post-processing \circled{8}. Finally, the post-processor synthesizes the final results and returns them to the user \circled{9}.

\begin{figure*}[t]
    \centering
    \includegraphics[width=0.8\textwidth]{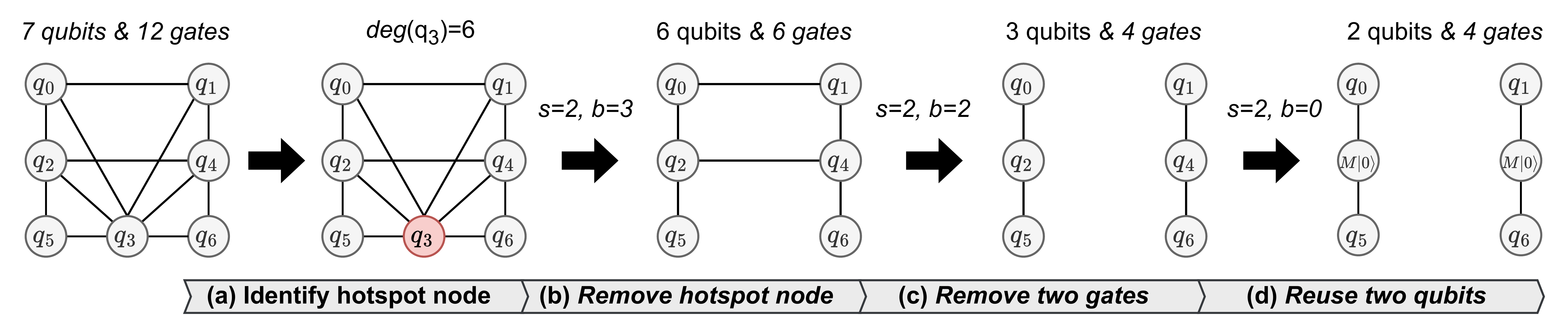}
    \caption{Error mitigator pre-processor workflow (\S~\ref{sec:error_mitigator:pre_processor}). {\em The initial Qernel has 7 qubits and 12 gates, and budget $b=3$. {\bf (a)} We identify $q_3$ as a hotspot node with a degree of 6. {\bf (b)} We remove this node, which reduces the qubit and gate counts to 6. {\bf (c)} We remove two gates, giving two fragments of 3 qubits each. {\bf (d)} The budget is depleted, so we use qubit reuse to further reduce the qubits by two.}}
    \label{fig:qernel_optimizer}
\vspace{-3mm}
\end{figure*}

\section{Error Mitigator}
\label{sec:error-mitigator}

Quantum computers are characterized by hardware and environmental noise errors, which hinder their practicality and scalability (\S~\ref{sec:design-challenges}). The error mitigator applies pre- and post-processing to the Qernels and execution results, respectively, to mitigate these noise errors.

\myparagraph{Challenges}
Currently, there is a plethora of individual error mitigation techniques that require their own sub-systems to operate, with no common infrastructure to compose them. \coloredText{However, composability is essential for fidelity improvement since error mitigation techniques can complement each other, if stacked together in the correct order, to further improve performance \cite{brenner2023optimal, brandhofer2023optimal}.}  Additionally, some techniques spawn an exponential number of sub-circuits and, after execution, require post-processing using classical resources \cite{peng2020simulating, mitarai2021constructing}. There are two main challenges in efficiently leveraging error mitigation: (1) Which error mitigation techniques must be used and \coloredText{in which order}? (2) How to balance the tradeoff between fidelity improvement and exponential overheads w.r.t. runtime and resources?

\myparagraph{Key ideas}
The error mitigator automatically applies error mitigation techniques, abstracting this complexity away from the user. The key idea is that not all gates and qubits are equal; some of them are noise \textit{hotspots}, lowering the fidelity significantly more than others. We constrain the exponential overheads by (1) setting an error mitigation budget $b$ to be spent, and (2) greedily applying error mitigation techniques on the hotspots \coloredText{in the order that maximizes fidelity improvement the most with the least added overheads}. 

\subsection{Pre-Processor}
\label{sec:error_mitigator:pre_processor}

The pre-processor first analyzes the Qernel to generate important metadata, identifies the error mitigation opportunities, and then prepares the Qernel for the respective techniques.

\myparagraph{Qernel analysis}
Qernel analysis generates two main pieces of information that guide the application of error mitigation techniques: (1) Qernel metadata and (2) optimization opportunities such as \textit{hotspot} qubits or gates, i.e., qubits and gates that can be removed to reduce noise errors significantly. Specifically, we generate the key Qernel static properties and the six SupermarQ feature vectors as stated in \S~\ref{sec:overview:programming_model}.

\myparagraph{Error mitigation techniques}
We focus on two \coloredText{existing} error mitigation techniques: circuit divide-and-conquer and qubit reuse, since these techniques increase fidelity and enable the execution of large quantum jobs on small(er) QPUs. \coloredText{However, \projecttitle{} supports any technique that achieves the same goals \cite{tannu2019mitigating, bravyi2021mitigating, Maciejewski_2020, das2021jigsaw}}.

At a high level, techniques in the former category \coloredText{divide} (large) circuits into smaller fragments that can be executed on small QPUs, and the fragmented execution results are \coloredText{post-processed} back to a single value. \coloredText{The advantages are two-fold: (1) We can scale the size of circuits that are executable on the current small QPUs beyond the H/W qubit limit, and (2) we can improve fidelity since smaller and less complex circuits achieve higher fidelity (\S~\ref{sec:background:foundations}, \S~\ref{sec:challenges:scalability}).} Unfortunately, these techniques incur \textit{exponential} quantum and/or classical overheads \coloredText{with respect to the number of divisions}; therefore, they must be used conservatively.

\coloredText{More specifically, circuit divide-and-conquer comprises \textit{circuit cutting and knitting} \cite{peng2020simulating, mitarai2021constructing} and \textit{qubit freezing} \cite{Ayanzadeh2023frozenQbits}. Circuit cutting can simplify the circuit's structure by virtualizing noisy non-local gates into (less noisy) local gates \cite{mitarai2021constructing} or cutting a qubit (wire) in the temporal dimension to shorten its runtime \cite{peng2020simulating}. Thus, circuit cutting can improve the circuit's width, depth, and number of non-local gates at once (\S~\ref{sec:background:foundations}). During knitting, the original (full) circuit results are reconstructed through classical post-processing of the subcircuit results. We refer interested readers to CutQC \cite{tang2021cutqc} and QVM \cite{tornow2024scaling} for details on how circuit cutting works in practice.  
Comparably, qubit freezing identifies qubits with significantly more connections to other qubits (hotspots) and partitions the circuit by replacing the hotspot qubits with binary values, effectively dropping their noisy non-local gates. Since every qubit can be replaced by a binary value, this process generates $2^m$ smaller circuits for $m$ frozen qubits.
}

 In both cases, we automatically find the cut locations that achieve the smallest sub-circuits possible with the fewest cuts possible. \coloredText{To maximize the synergy between techniques, we first apply qubit freezing to remove multiple non-local gates at once with a relatively small cost, and then we greedily apply circuit cutting.}
 To restrict the exponential overheads that scale exponentially with the number of cuts, we use the budget $b$ to cut up to $b$ times. Typically, we set the budget $b=3$ by default, and the overheads scale as $O(2^b)-O(8^b)$, depending on the divide-and-conquer technique.

\coloredText{On the other hand, qubit reuse (also referred to as recycling) reduces qubit requirements by reusing physical qubits after resetting them \cite{hua2023caqr, decross2022qubitreuse, jiang2024qubit}. Once a qubit's role in computation is complete, it is measured, reset to a known state (typically $\ket{0}$), and reassigned to another logical qubit later in the circuit. This approach enables more efficient use of limited hardware resources and allows execution of larger circuits than the QPU's qubit limit.}
This method, however, increases the \coloredText{circuit depth (i.e., execution duration)}, which can lead to quantum decoherence errors (\S~\ref{sec:background:foundations}); therefore, the tradeoff, in this case, is between circuit size and runtime. To restrict the runtime increase, we use qubit reuse as a last resort to render the Qernel executable by at least one QPU in the system \coloredText{(i.e., the optimized Qernel fits at least one QPU's size constraints)}.

\myparagraph{Workflow example}
Figure \ref{fig:qernel_optimizer} shows the pre-processing workflow for a QAOA circuit (\S~\ref{sec:background:101}) with 7 qubits and 12 gates. The pre-processor aims to achieve a maximum Qernel size of two qubits with a budget of $b=3$. To achieve this, it takes the following steps:

\noindent
\textbf{Step 1:} 
The pre-processor applies analysis to identify a hotspot node, in this case, $q_3$ is a hotspot with a degree of 6 gates (Figure \ref{fig:qernel_optimizer}, (a)).

\noindent
\textbf{Step 2:} 
Then, the pre-processor removes $q_3$ and its gates \coloredText{with qubit freezing}. The new Qernel size is 6 qubits with 6 gates. Then, it updates the budget to $b=b-m$, where $m=1$ is the number of qubits frozen (Figure \ref{fig:qernel_optimizer}, (b)).

\noindent
\textbf{Step 3:} 
Next, the pre-processor applies circuit cutting on two gates and updates the budget to $b=0$. The new Qernel consists of two fragments, each with a size of 3 qubits and 4 gates (Figure \ref{fig:qernel_optimizer}, (c)). 

\noindent
\textbf{Step 4:} 
Since $b=0$, the pre-processor applies qubit reuse to achieve a size of 2 qubits and identifies qubits $q_0,q_1$ as reusable. \coloredText{In the Figure, $M\ket{0}$ denotes measurement and reset to the $\ket{0}$ state}. The final Qernel now has two fragments of 2 qubits and 4 gates (Figure \ref{fig:qernel_optimizer}, (d)). 

The final output is a Qernel with a 42.8\% smaller size and 66\% less noisy gates. \coloredText{Notably, this result is only attainable through the synergy of techniques, as no individual method alone can achieve comparable improvements.}

\subsection{Post-Processor}
The post-processor \coloredText{reconstructs} the final error-mitigated results by \coloredText{classically stitching together intermediate outcomes from sub-Qernel executions. In circuit knitting, each virtualized gate (i.e., cut) is expanded into a linear combination of basis gates with associated coefficients. When multiple such gates are virtualized, the global coefficient vector becomes the tensor product of the individual gate vectors, resulting in an exponentially large space of up to $8^k$ bitstring-weight pairs for $k$ virtualized gates. The post-processing then requires computing a weighted sum over all $8^k$ combinations, with each term involving a product of subcircuit results. As such, this process requires a scalable post-processing infrastructure.}

\myparagraph{The map phase}
To efficiently process the large number of results, we follow a divide-and-conquer approach. Specifically, we split the results into $k$ equal sizes and distribute them to $k$ classical nodes to be processed in parallel. We parallelize across $k$ to increase data locality and reduce communication overheads since all results for each of the $k$ cuts will be in the memory of the same node.  Locally, each node performs tensor product ($\bigotimes$) operations on the probability distributions, which are parallelizable across the node's threads. If available in the node, \projecttitle{} leverages GPUs or TPUs to accelerate the tensor products. Following this process, the $k$ nodes output $k$ intermediate results, ready to be reduced into a single result. 

\myparagraph{The reduce phase}
\projecttitle{} selects any of the $k$ nodes to perform the reduce step. The rest of the nodes send the intermediate results to this node, which performs a thread-parallel sum of $k$ results. Equivalent to the map phase, the parallel sum can also be executed on GPUs. This produces the final output to be returned to the user.

\section{Estimator}
\label{sec:qos:estimator}

The estimator is responsible for predicting the fidelity performance of a given Qernel on the underlying QPUs without executing the Qernel, which would be extremely costly. This prediction will be the leading decision factor for the scheduler when assigning the Qernel to a QPU. 

\myparagraph{Challenges}
Estimating fidelity at-priori faces three key challenges: (1) Fidelity is a non-deterministic metric affected by the hardware's probabilities of errors, which change across QPUs and time unpredictably (\S~\ref{sec:design-challenges}). (2) Simulating a Qernel to obtain fidelity estimates is intractable since the complexity of simulating quantum systems scales exponentially with the number of qubits \cite{georgescu2014quantum}. (3) Fidelity can be approximated numerically \cite{mapomatic}; however, there is an accuracy-cost tradeoff. The more complex the analytical model becomes, the higher the runtime overhead of estimation.

\myparagraph{Key idea}
The estimator leverages analytical \coloredText{and regression} models that do not require real execution or simulation of the Qernel. Specifically, the estimator's models compute a \textit{score} for each Qernel-QPU assignment that captures the potential fidelity of that assignment, and the estimator supports configurable scoring policies with different runtime-cost tradeoffs. These policies consider (1) the Qernels' properties generated from the error mitigator (2) the QPUs' calibration data \coloredText{(i.e., noise model)}, which are available to quantum cloud providers since they perform the calibration cycles.

For (1), important properties include the number and types of gates, depth, and the number of measurements (\S~\ref{sec:qernel:properties}). For (2), recall that QPUs are characterized by calibration data that describe the exact error rates of the QPU for that calibration cycle (\S~\ref{sec:background:foundations}), specifically, the individual qubit readout errors, the individual gate errors, and the $T2$ coherence times. In this work, we implement two scoring policies: a numerical approach for fine-grained control over the estimations and a regression model approach for abstracting away the complexity of estimation.

\myparagraph{Numerical cost policy} 
This policy estimates execution fidelity by transpiling the circuit for the target QPU (\S~\ref{sec:background:foundations}). Target transpilation enables fine-grained fidelity estimation by producing the mapping between logical and physical qubits and the gate (instruction) schedule. The mapping captures the expected readout and gate errors, while the gate schedule captures the order and exact timing that the gates will be applied on the qubits, which reveals the hardware decoherence and crosstalk errors, as explained in \S~\ref{sec:background:foundations}. \coloredText{This method has been explored before, and there are multiple variants, but we base our implementation on Mapomatic \cite{mapomatic}.} 

\coloredText{To compute the score, we need the following information: (1) The readout/measurement error probabilities that describe the probability of a bit-flip during measurement, (2) the gate error probabilities, (3) the T2 time, which measures for how long a qubit can stay in an excited state (i.e., $\ket{1}$), thus quantifying the decoherence error probability, and (4) the crosstalk error which is measured through crosstalk characterization \cite{murali2020software}. With this information, we can multiply the probabilities of the respective sources of errors to get an estimate of the total error/fidelity.}

Formally, for each qubit $q_i$ the readout error is $e_{r(i)}$, for each gate $g_j$ the error is $e_{g(j)}$, and the decoherence error is $e_{d(t)} = 1 - e^{-t/T2_i}$, where $t$ is the idle time of the qubit $q_i$ (no gates act on it \cite{das2021adapt}) and $T2$ is the decoherence time of $q_i$. The crosstalk error between gates $g_k$ and $g_l$ is $e_{ct(k,l)}$. Putting it all together, the final fidelity score is computed as follows: $fid=1- \prod_{i=0}^{N} e_{r(i)}e_{d(i)} \prod_{j=0}^{M} e_{g(j)} \prod_{j=0, k=0}^{M\times M}e_{ct(j,k)}$, where $N$ is the circuit's number of qubits and $M$ is the number of gates. 
Since all hardware error information is known at-priori, and quantum errors accumulate multiplicatively, this policy produces high-accuracy estimations, as we show in \S~\ref{evaluation:results:estimator}.  

\begin{figure*}[ht]
    \centering
    \includegraphics[width=0.9\textwidth]{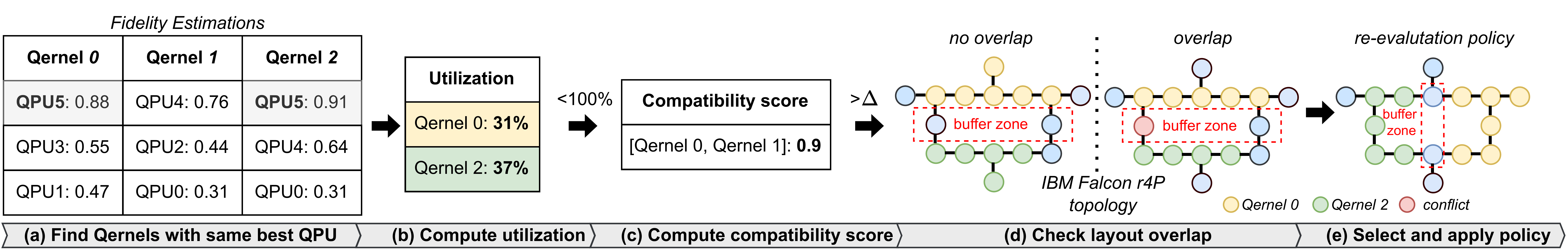}
    \caption{\projecttitle{} multi-programmer example workflow (\S~\ref{sec:qos:multi-programmer}). {\em {\bf (a)} We use the estimator's output to find Qernels with the same best QPU. {\bf (b)} We compute their independent utilization, and {\bf (c)} their compatibility score. If compatible, {\bf (d)} we check for layout overlap, and {\bf (e)} apply the appropriate multi-programming policy.}}
    \label{fig:multi-programmer}
\vspace{-3mm}
\end{figure*}

\myparagraph{Regression model policy}
Since the impact of noise errors on fidelity during quantum computation can be described \coloredText{numerically}, we can train a regression model to predict the fidelity of a transpiled Qernel on a possible QPU using the QPU's calibration data and the Qernel's static properties as features.  Specifically, we use the aforementioned errors we defined in the numerical cost policy as QPU features and the static properties (\S~\ref{sec:qernel:properties}) as Qernel features. Even simple regression models such as linear regression achieve high prediction accuracy, up to 99\%. Although this policy is simple to use without detailed knowledge of the relationship between errors, in \projecttitle{}, we use the numerical cost policy by default for estimation to have a clear understanding and full control of the process.

\section{Multi-programmer}
\label{sec:qos:multi-programmer}

The size of quantum programs that run with high fidelity is small, leading to QPU underutilization (\S~\ref{sec:challenges:utilization}). To increase QPU utilization, \projecttitle{} \textit{multi-programs} two or more Qernels, potentially from different users, to run on the same QPU. We refer to this multi-programming as \textit{bundling} the Qernels together.

\myparagraph{Challenges}
Bundling Qernels faces three distinct challenges: (1) Trivially bundling Qernels together will deteriorate fidelity because qubits interfere with each other via crosstalk errors (\S~\ref{sec:background:foundations}). (2) On top of that, bundled Qernels with unequal runtimes do not optimally increase utilization since \textit{effective} utilization is measured in space and time (number of QPU qubits used and duration that they are non-idle). (3) Qernel compilation involves several NP-hard processes \cite{siraichi2018qubit}, so we need to minimize it. 

\myparagraph{Key ideas}
To tackle these challenges, we \coloredText{introduce new ideas unexplored by prior multi-programming work \cite{das2019a, liu2021qucloud}}. First, we define a new utilization metric, namely \textit{effective} utilization, that captures the \coloredText{temporal dimension of utilization, i.e., H/W qubit usage during computation time}. Next, we compute Qernel \textit{compatibility functions} that quantify how well-suited any two Qernels are to run together, \coloredText{to minimize fidelity loss and re-compilation times}. Last, we create \textit{buffer zones} between two bundled Qernels to minimize crosstalk \coloredText{errors}.

\subsection{Qernel Compatibility}
Qernel compatibility quantifies crosstalk errors and the effective utilization of bundled Qernels by considering the Qernels' static properties (\S~\ref{sec:qernel:properties}).

\myparagraph{Effective utilization}
\coloredText{The trivial way to compute utilization in the case of bundled Qernels is by diving their total number of qubits over the QPU's number of qubits, $N_{total}/N_{QPU}$. However, this is not accurate when the bundled Qernels have unequal runtimes.}
\coloredText{In order to accurately quantify QPU usage in the context of multi-programming, we define effective utilization as \textit{spatial} plus \textit{temporal} utilization.}
\coloredText{To quantify spatial utilization, it suffices to compute the ratio of allocated QPU qubits (of the longest Qernel in terms of depth) over the QPU's number of qubits}. 

\coloredText{To quantify temporal utilization, instead of simply adding the spatial utilization of each allocated Qernel to the total utilization, we first scale it with a weight that is the ratio of the Qernel's duration over the longest Qernel's duration. Recall that, to measure the Qernels' durations, we use the depth property that reflects the longest sequence of gates the Qernel consists of (\S~\ref{sec:background:foundations}).}

More technically, we define effective utilization as $u_{eff}=\frac{N_{C_{max}}}{N_{QPU}} * 100 + \sum_{k=1}^{n} \frac{D_k}{D_{max}} * \frac{N_{C_k}}{N_{QPU}} * 100$, where $N_{C_{max}}, N_{QPU}$ are the number of qubits of the longest Qernel and the QPU, respectively,  $k$ is the number bundled Qernels excluding the longest Qernel, and $D$ is the depth of the Qernel. \coloredText{The first term, $\frac{N_{C_{max}}}{N_{QPU}} * 100$, captures the spatial utilization. The right term (sum) captures the temporal dimension, and $\frac{D_n}{D_{max}}$ is the relative depth of Qernel $k$ compared to the longest Qernel, and $\frac{N_{C_k}}{N_{QPU}}$ is the fraction of QPU qubits used by Qernel $k$.}

For example, a 10-qubit Qernel $Q_0$ spatially utilizes 50\% of a 20-qubit QPU. Now assume that $Q_0$ runs 3$\times$ longer than a 10-qubit Qernel $Q_1$. During $\frac{2}{3}$ of $Q_0$'s runtime, the qubits allocated to $Q_1$ will be idle, decreasing the effective utilization to only 66\%.

\myparagraph{Quantifying crosstalk}
To quantify crosstalk without running the bundled Qernels, we use the \textit{entanglement ratio} and \textit{parallelism} Qernel feature vectors of \cite{tomesh2022supermarq}, where higher values indicate a higher chance for crosstalk errors (\S~\ref{sec:background:foundations}). Intuitively, the entanglement ratio captures the proportion of 2-qubit gates over all gates, and parallelism captures how many gates run in parallel per time unit. Thus, more parallel 2-qubit gates equal a higher chance of crosstalk errors \cite{murali2020software}.

\myparagraph{Compatibility formula}
\coloredText{To quantify the compatibility of two candidate Qernels, we consider their effective utilization, as well as their joint entanglement ratio and parallelism scores. This is because we ideally want high effective utilization (not just higher spatial) and fewer crosstalk-induced errors.}

Formally, we score a possible Qernel pair as follows: $qc=\alpha~ u_{eff} + \beta ~ER_{b} + \gamma~ PA_{b} \mapsto [0,1]$. Higher score is better, $\alpha + \beta + \gamma = 1$ \coloredText{(the weights add up to one), $u_{eff}$ is effective utilization},  $ER$ is entanglement ratio, $PA$ is parallelism, and $b$ denotes \textit{bundled}, i.e., $E_b$ is the entanglement ratio of the bundled Qernels. The four variables are tunable to give priorities on different objectives, e.g., prioritize effective utilization or minimize crosstalk. After experimenting and fine-tuning, we found that $\alpha=0.25, \beta=0.25, \gamma=0.5$, and $qc \geq 0.75$ gives balanced results, as we show in \S~\ref{evaluation:results:multi-programmer}.

\myparagraph{Buffer zone}
Crosstalk characterization studies on real machines have shown that the probability of crosstalk errors drops exponentially with the distance between any two qubits \cite{murali2020software}. We leverage this \coloredText{to minimize crosstalk errors} by creating a \textit{buffer zone} between any two bundled Qernels, i.e., at least one hardware qubit between the two sets of allocated qubits must be free. Given the hardware constraints and limited number of available qubits, we limit the buffer size to up to two qubits of distance.

\subsection{Multi-programming Policies}

In this work, we implement two multi-programming policies; the first is the \textit{fast path} multi-programming while the second requires re-compilation and re-estimation.

\myparagraph{Restrict policy}
The restrict policy checks if there is no overlap in the Qernel mappings on the QPU, including the buffer zone. For example, in Figure \ref{fig:multi-programmer} (d) in the left case, the logical qubits of the Qernels are mapped to disjoint sets of physical qubits on the QPU, and there is a buffer zone between the two sets. In that case, the policy simply bundles the Qernels together, and fidelity loss is minimized through the aforementioned compatibility score.

\myparagraph{Re-evaluation policy}
This policy is the fallback of the restrict policy. If the Qernel mappings overlap, the two Qernels are transpiled again for the target QPU, and their new fidelity is estimated. If the new fidelity is lower up to a fixed $\epsilon > 0$ value compared to the original independent fidelities, the bundling is kept. Otherwise, the multi-programmer selects the next most compatible Qernel pair.

\myparagraph{Example}
Figure \ref{fig:multi-programmer} shows an example where the multi-programmer receives three Qernels with three estimations each and identifies $Qernel~0$ and $Qernel~2$ as a possible pair since their best QPU is the same ($QPU5$) (a). It computes their independent utilization (31\% and 37\%, respectively), and the combined utilization is under 100\% (b). It computes the compatibility score that surpasses the threshold ($0.9>0.75$) (c). In (d), the right case shows an overlap in the buffer zone (the red qubit), so we apply the re-evaluation policy.

\myparagraph{Unbundling}
To unbundle the results, the multi-programmer keeps a record that maps the initial (solo) Qernel IDs to the new, bundled Qernel ID, as well as the Qernels' sizes. Therefore, when receiving a new result from a Qernel with an ID $i$, it scans the record to find an entry $i$, and if found, it splits the probability distribution bitstrings  (\S~\ref{sec:background:101}) into two parts: the left-most and the right-most bits based on the Qernel sizes.

\section{Scheduler}
\label{sec:qos:scheduler}

\begin{figure*}[ht] 
    \centering
    \includegraphics[width=\textwidth]{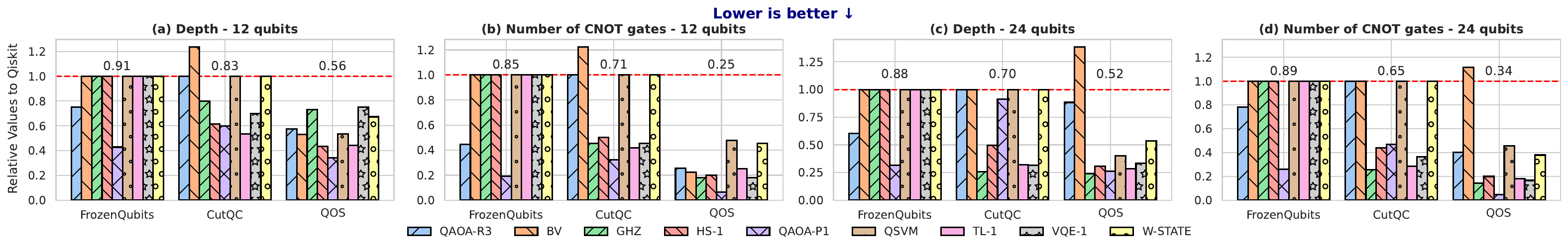}
   \caption{Error mitigator (\S~\ref{sec:evaluation:compiler}). {\em Impact of the error mitigator on the circuit depth and the number of CNOTs. The circuits are optimized using budget $b=3$, and we compare against Qiskit (red horizontal line), FrozenQubits \cite{Ayanzadeh2023frozenQbits} and CutQC \cite{tang2021cutqc}.
    There is an average $46\%$, $38.6\%$, and $29.4\%$ reduction in circuit depth, and  $70.5\%$, $66\%$ and $56.6\%$ reduction in the number of CNOTs, respectively.}}   
    \label{fig:eval:dt:relative_properties}
    \vspace{-4mm}
\end{figure*}

Scheduling quantum programs involves fundamental tradeoffs between conflicting objectives; specifically, users ideally want the fidelity of \textit{the best fidelity QPU} and the waiting time of \textit{least busy QPU}.

\myparagraph{Challenges}
Scheduling quantum tasks faces three key challenges: (1) To maximize fidelity, most programs must be scheduled on the same subset of best-performing QPUs (\S~\ref{sec:challenges:spatial_and_temporal}), forming hotspots and necessarily increasing user waiting times. (2) The scheduler must at least know estimates of the quantum task durations. However, similar to fidelity estimation (\S~\ref{sec:qos:estimator}), quantum execution time estimation must be fast, without real execution or simulation. (3) Job scheduling is a well-known NP-hard problem; therefore, every heuristic or greedy solution will present a tradeoff between optimality and performance.  

\myparagraph{Key ideas}
Our scheduler leverages an analytical model to estimate the Qernel execution time quickly. Then, based on the fidelity predictions from the estimator, it assigns and runs Qernels across space (which QPUs) and time (when). Last, our scheduler supports configurable policies for managing the aforementioned tradeoffs, prioritizing maximal fidelity, minimal waiting times, or a balanced approach. Our two main policies are a fast heuristic algorithm and a genetic multi-objective optimization algorithm.

\myparagraph{Execution time estimation}
To optimize for minimal waiting times, the scheduler must first estimate each Qernel's execution time and then aggregate the execution time estimations in each QPU's queue to compute the total waiting times. 
To estimate the execution time, we iterate the longest path (depth) of a Qernel (\S~\ref{sec:qernel}) that corresponds to the longest-duration gate chain and thus defines the Qernel's execution time. By summing the gate durations of each node in the longest path, we get the Qernel's total execution time.

\myparagraph{Formula-based policy}
Optimizing for conflicting objectives involves comparing two possible solutions (e.g., maximal fidelity vs. minimal waiting times). In the formula-based policy, we use the following scoring formula: Score = $c\frac{f_2-f_1}{f_1} - (1-c)\frac{t_2-t_1}{t_1} + \beta \frac{u_2-u_2}{u_1}$ to compare and select between two possible assignments. This formula factors fidelity, waiting time, and utilization to determine which assignment is better, given priorities. The parameters are as follows: \textbf{$f_i$}: fidelity of the estimation result $i$, \textbf{$t_i$}: waiting time for the QPU from estimation result $i$, \textbf{$u_i$}: utilization of the QPU for estimation result $i$, \textbf{$c\in(0,1)$}: a system-defined constant that weighs the fidelity difference between estimations and finally, \textbf{$\beta$}: a system-defined constant acting as a weighting factor for utilization difference, balancing system throughput and fidelity. By selecting higher $c$, the system prioritizes fidelity over waiting times, and vice versa, and by selecting higher $\beta$ the system prioritizes utilization over fidelity, and vice versa. By default, $c=\beta=0.5$, which aims for balanced fidelity, waiting times, and utilization.

\myparagraph{Genetic algorithm policy}
Genetic algorithms excel at optimizing for conflicting objectives by efficiently searching over vast search spaces, and for that, they can be used in the context of \projecttitle{}. We formulate a multi-objective optimization problem with the conflicting objectives of fidelity vs. waiting times and use the NSGA-II genetic algorithm \cite{deb2002a} to solve it. The algorithm creates a Pareto front of possible solutions (schedules), each achieving a different combination of average fidelity and average waiting times. Then, to select one of those schedules, we use the aforementioned formula to score each schedule and select the schedule with the highest score.

\section{Evaluation}
\subsection{Experimental Methodology}
\label{sec:evaluation:experimental_methodology}

\myparagraph{Experimental setup}
We conduct two types of experiments: (1) classical tasks, such as circuit transpilation and trace-based simulations, and quantum tasks (2), which run on \textbf{\textit{real}} QPUs for measuring the circuits' fidelities.

\begin{figure*}[ht]
    \centering
    \includegraphics[width=0.95\textwidth]{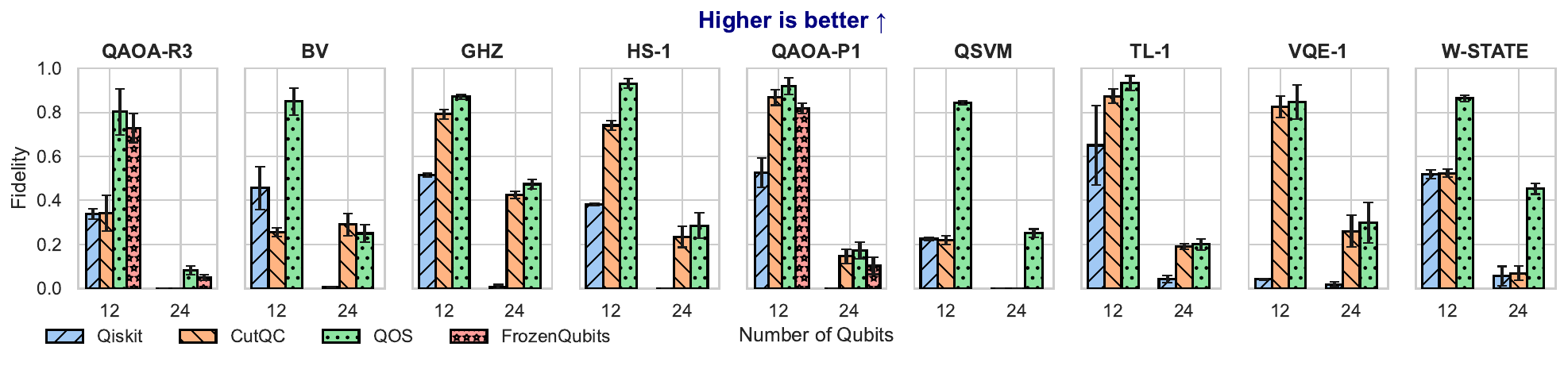}
    \caption{Error mitigator (\S~\ref{sec:evaluation:compiler}). {\em Impact of the error mitigator on the circuit fidelity against Qiskit \cite{qiskit-transpiler}, CutQC \cite{tang2021cutqc}, and FrozenQubits \cite{Ayanzadeh2023frozenQbits}. The circuits are optimized using budget $b=3$. There is a mean $2.6\times$, $1.6\times$, and $1.11\times$ improvement for 12-qubit circuits, respectively.  
    There is a $456.5\times$, $7.6\times$, and $1.67\times$ improvement for circuits of 24 qubits, respectively.}}
    \label{fig:eval:dt:fidelities}
    \vspace{-5mm}
\end{figure*}

For (1), we use a server with a 64-core AMD EPYC 7713P processor and 512 GB ECC memory.
For (2), we conduct our experiments on IBM Falcon r5.11 QPUs. Unless otherwise noted, we use the \textbf{\textit{real}} Kolkata 27-qubit QPU hosted by IBM.

\myparagraph{Framework and configuration} We use the \textit{Qiskit} \cite{Qiskit} Python SDK for compiling quantum circuits and running simulations. We compile quantum circuits with the highest optimization level (3) and run with 8192 shots. Each data point presented in the figures is the median of five runs.

\myparagraph{Benchmarks}
We study \projecttitle{} on a set of circuits used in state-of-the-art NISQ algorithms, adopted from the 3 benchmark suits of Supermarq \cite{tomesh2022supermarq}, MQT-Bench \cite{quetschlich2022mqt} and QASM-Bench \cite{li2023qasmbench}. The algorithms' circuits can be scaled by the number of qubits and depth. Specifically. We study 9 benchmarks: \textup{GHZ}, \textup{W-State}, \textup{Bernstein Vazirani (BV)}, \textup{Hamiltonian Simulation (HS-$t$)}, \textup{Quantum-enhanced Support Vector Machine (QSVM)}, \textup{Two Local Ansatz (TL-$n$)}, \textup{Variational Quantum Eigensolver (VQE-$n$)}, and \textup{Approximate Optimization Algorithm (QAOA-R/P)}, these benchmarks cover a wide range of relevant criteria for evaluating \projecttitle{}.

For the TL and VQE circuits, we use circular and linear entanglement, respectively.
The HS, VQE, and TL benchmarks are scalable by their circuit depth with the number of time-steps $t$ and layers in the $ansatz$ $n$. The QAOA-R/P circuits are initialized using regular/power-law graphs, respectively, with degree $d \in \{1, 3\}$.

\myparagraph{Metrics} We evaluate the following metrics: {\bf (1) Hellinger Fidelity}. As defined in \cite{fidelity-qiskit}, where it ranges in $[0,1]$ and higher is better. {\bf (2) Static Properties}. Number of \textit{CNOT} gates and \textit{depth} (\S~\ref{sec:background:foundations}). {\bf (3) Waiting Time}. The time a Qernel spends in a QPU's queue, waiting for execution, in seconds. {\bf (4) Classical Overhead}. The error mitigation classical overheads as a factor of runtime increase ($\times$). {\bf (5) Quantum Overhead}. The error mitigation quantum overheads as a factor of numbers of circuits increase ($\times$).

\myparagraph{Baselines}
We evaluate the error mitigator against \coloredText{the standalone fidelity-improving frameworks} Qiskit v0.41, CutQC \cite{tang2021cutqc} and FrozenQubits \cite{Ayanzadeh2023frozenQbits} \coloredText{to show the efficacy of the error mitigator's composability}. We evaluate \projecttitle{}'s multi-programmer against \cite{das2019a}, \coloredText{the state-of-the-art multi-programming framework}. Regarding \projecttitle{} scheduler, to the best of your knowledge, \cite{ravi2021adaptive} is the only peer-reviewed quantum scheduler, but it doesn't provide source code or enough technical details to faithfully implement it. 
\coloredText{Lastly, we do not perform end-to-end evaluation against a unified baseline since simply combining prior works into a pipeline does not capture the cross- and within-layer optimizations that \projecttitle{} performs.}

\subsection{\projecttitle{} Error Mitigator}
\label{sec:evaluation:compiler}

\begin{figure*}[t]
    \centering
    \includegraphics[width=0.95\textwidth]{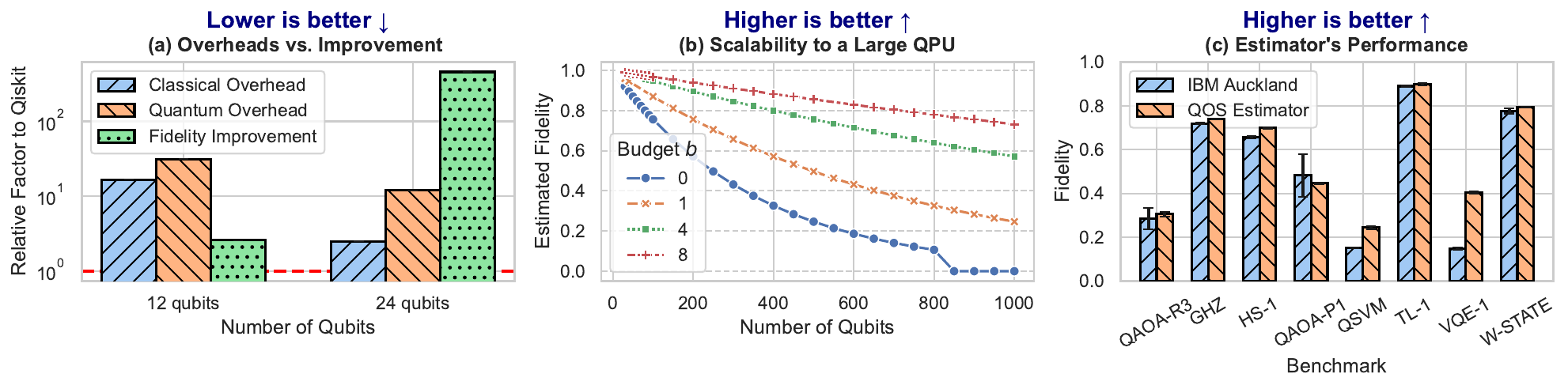}
    \vspace{-3mm}
    \caption{Error mitigator (\S~\ref{sec:evaluation:compiler}) and estimator (\S\ref{evaluation:results:estimator}). {\em {\bf (a)} Error mitigator: classical and quantum overheads and fidelity improvement as a relative factor to Qiskit. For 24 qubits, the improvement outweighs the overheads. {\bf (b)} Error mitigator: scalability to a large, hypothetical 1000-qubit QPU. Any budget $b>0$ achieves higher quality results than using no optimizations. {\bf (c)} Estimator's performance: fidelity of IBM Auckland vs. the QPU automatically selected by the estimator.}}
    \label{fig:compiler_overheads}
    \vspace{-3mm}
\end{figure*}

{\bf RQ1:} \textit{How well does the error mitigator improve the fidelity of circuits that run on NISQ QPUs?}
We evaluate the performance of the error mitigator w.r.t the post-mitigation properties and fidelity of the circuits while also analyzing the classical and quantum costs of our approach.

\myparagraph{Effect on the circuit depth and number of CNOTs}
In Figure \ref{fig:eval:dt:relative_properties}, we show the performance of the error mitigator on the circuits' depth and number of CNOTs, where we plot the relative difference in post-optimization circuit depth and the number of CNOTs between Qiskit (the red horizontal line) and FrozenQubits \cite{Ayanzadeh2023frozenQbits}, CutQC \cite{tang2021cutqc}, and our error mitigator. Figures \ref{fig:eval:dt:relative_properties} (a) and (c) show that the circuit depth decreases by $46\%$, $38.6\%$, and $29.4\%$, respectively. Figures \ref{fig:eval:dt:relative_properties} (b) and (d) show that the number of CNOTs decreases by $70.5\%$, $66\%$, and $56.6\%$, respectively.
The improvement in both metrics against the baselines is attributed to the composability of our error mitigator, where the combined techniques achieve better results than any standalone technique. \coloredText{In cases where \projecttitle{} under-performs compared to the baselines  (e.g., BV benchmark for 24 qubits), it is because the error mitigator achieved the desired circuit size with the qubit reuse technique, which incurs additional costs (\S\ref{sec:error_mitigator:pre_processor}).}

\myparagraph{Impact on fidelity}
Figure \ref{fig:eval:dt:fidelities} shows the error-mitigated circuits' fidelity against Qiskit \cite{qiskit-transpiler}, CutQC \cite{tang2021cutqc}, and FrozenQubits \cite{Ayanzadeh2023frozenQbits}. The results show a mean $2.6\times$, $1.6\times$, and $1.11\times$ improvement for 12-qubit circuits, respectively, and a $456.5\times$, $7.6\times$, and $1.67\times$ improvement for circuits of 24 qubits, respectively. The fidelity improvement is a consequence of lower circuit depths and fewer CNOTs, as shown in
 Figure \ref{fig:eval:dt:relative_properties}. 

\myparagraph{Classical and quantum overheads}
Figure \ref{fig:compiler_overheads} (a) shows the average classical and quantum overheads of the error mitigator. The classical overhead is $16.6\times$ and $2.5\times$ for 12 and 24 qubits, respectively, and the quantum overhead is 31.3$\times$ and 12$\times$ for 12 and 24 qubits, respectively. However, fidelity improves by $2.6\times$ and $456.5\times$ for 12 and 24 qubits, respectively; therefore, for larger circuits, the fidelity improvement is worth the cost. 

\myparagraph{Scalability}
To demonstrate that the error mitigator is scalable, we run the VQE-1 benchmark on a hypothetical 1000-qubit QPU with one-qubit gate errors of $10^{-4}$, two-qubit gate errors of $10^{-3}$, and measurement errors of $10^{-2}$. We optimize with budget $b\in\{0,1,4,8\}$ and report the estimated fidelity. Figure \ref{fig:compiler_overheads} (b) shows that all budget $b$ values improve the estimated fidelity, \coloredText{where higher $b$ equals higher fidelity.}

\myparagraph{RQ1 takeaway}
The error mitigator improves the properties of quantum circuits by $51\%$ on average, leading to an improvement in fidelity of 2.6--456.5$\times$, while incurring justifiable classical and quantum overheads.

\subsection{Estimator}
\label{evaluation:results:estimator}

{\bf RQ2:} \textit{How well does \projecttitle{}'s estimator address spatial and temporal heterogeneities?} We evaluate the estimator's precision in selecting the top-performing QPU for each benchmark. We establish a baseline using the on-average best-performing machine every calibration day. On the day of the experiment, IBM Auckland was the best-performing machine (also with the highest number of pending jobs).

\myparagraph{Estimator's accuracy}
Figure \ref{fig:compiler_overheads} (c) shows the fidelity of the eight benchmarks when run on QPUs selected by the estimator versus when run on the IBM Auckland QPU. The QPU selected for the BV benchmark is Auckland; therefore, we omit this result. For the rest of the benchmarks, the IBM Sherbrooke and Brisbane QPUs were automatically selected. Interestingly, the fidelity is on par or even higher than IBM Auckland, except for only one benchmark, the QAOA-P1, \coloredText{possibly because of its unique, power-law connectivity structure}. 

\myparagraph{RQ2 takeaway}
\projecttitle{}'s estimator automatically identifies QPUs with higher fidelity than the current standard practice.

\subsection{Multi-programmer}
\label{evaluation:results:multi-programmer}

{\bf RQ3:} \textit{How well does \projecttitle{}'s multi-programmer increase QPU utilization with minimum fidelity penalties?} We evaluate the impact of the multi-programmer on the fidelity of co-scheduled circuits for certain utilization thresholds. 

\begin{figure*}[ht]
    \centering
    \includegraphics[width=0.9\textwidth]{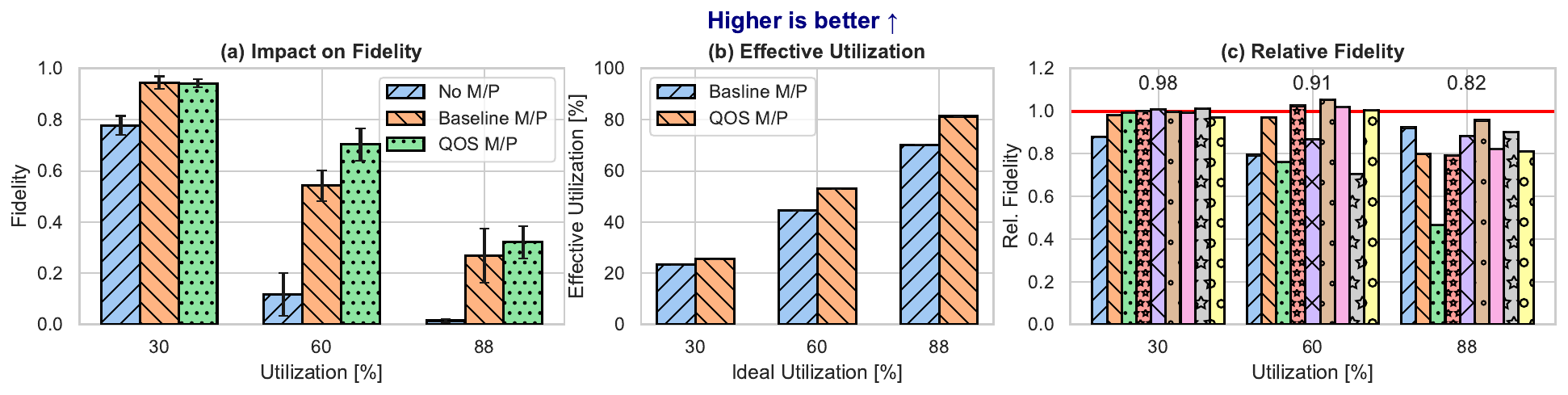}
    \caption{Multi-programmer (\S~\ref{evaluation:results:multi-programmer}). \textbf{(a)} {\em Impact of multi-programming on fidelity. There is a 9.6$\times$ and 1.15$\times$ improvement compared to no multi-programming and the baseline, respectively.} \textbf{(b)} {\em Effective utilization. There is 7.2\% higher effective utilization on average.} \textbf{(c)} {\em Relative fidelity w.r.t. solo circuit execution. There is an average 9.6\% drop in fidelity due to \projecttitle{}'s multi-programming.}}    
    \label{fig:eval:multi-programmer}
    \vspace{-3mm}
\end{figure*}

\myparagraph{Utilization vs. fidelity}
Figure \ref{fig:eval:multi-programmer} (a) shows the average fidelity of nine benchmarks with utilization of 30\%, 60\%, and 88\%. The three bars represent: no multi-programming \textit{(No M/P)} refers to large circuits that run solo, baseline multi-programming \textit{(Baseline M/P)} refers to \cite{das2019a}, and \projecttitle{}'s multi-programming approach \textit{(\projecttitle{} M/P)}. There is an average 9.6$\times$ improvement in fidelity compared to solo execution and an average 15\% (1.15$\times$) improvement compared to the baseline.

\myparagraph{Effective utilization}
The results in Figure \ref{fig:eval:multi-programmer} (b) show that \projecttitle{} achieves, on average, a 7.2\% higher effective utilization (\S~\ref{sec:qos:multi-programmer}), with a maximum improvement of 10.1\%. \coloredText{We attribute this improvement to the inclusion of temporal utilization.}

\myparagraph{Fidelity penalty vs. solo execution}
In Figure \ref{fig:eval:multi-programmer} (c), we evaluate the fidelity penalty of multiprogramming vs. solo circuit execution for utilization of 30\%, 60\%, and 88\%. The fidelity loss is 2\%, 9\%, and 18\%, respectively. The average fidelity loss is 9.6\% compared to solo execution, which is in line with previous studies \cite{das2019a, liu2021qucloud}. In the worst case (18\%), the fidelity loss is caused by the restrictions in high-quality qubit allocations and the crosstalk errors.

\begin{figure*}[ht]
    \centering
    \includegraphics[width=0.9\textwidth]{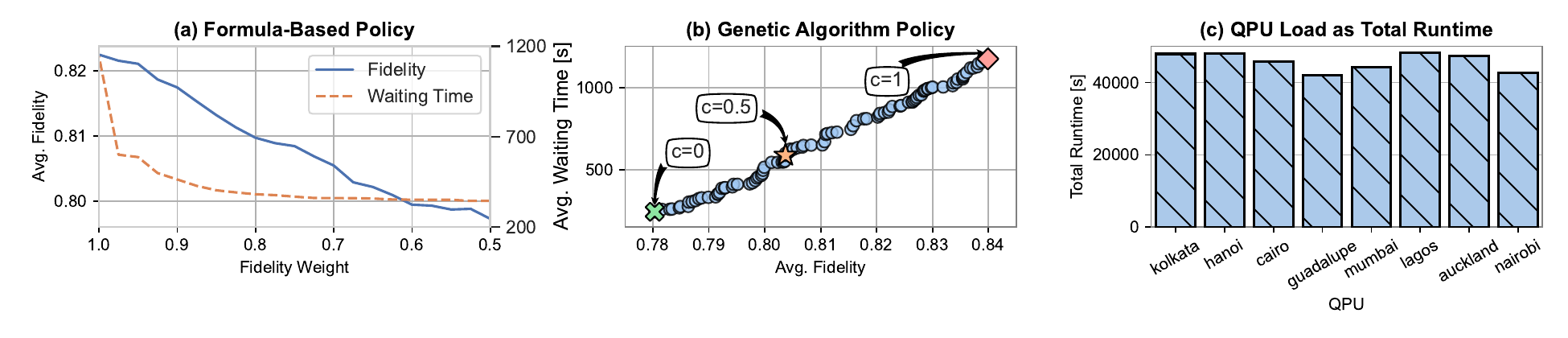}
    \vspace{-5mm}
    \caption{Scheduler (\S~\ref{evaluation:results:scheduler}). { \textbf{(a)} \em Formula-based scheduling policy: Average fidelity vs. average waiting times. A fidelity weight $c=0.7$ achieves $\sim5\times$ lower waiting time for only $\sim2\%$ lower fidelity}. {\textbf{(b)} \em Genetic algorithm policy: it creates a Pareto front of schedules, where a fidelity weight $c=0.5$ achieves $2\times$ lower waiting times for $\sim4\%$ fidelity decrease.} {\textbf{(c)} QPU load as the total runtime of each QPU for the formula-based policy. The maximum load difference between any two QPUs is $15.2\%$.}}
    \label{fig:avg_merged}
    \vspace{-3mm}
\end{figure*}

\myparagraph{RQ3 takeaway}
The \projecttitle{} multi-programmer improves fidelity by 1.15--9.6$\times$ and effective utilization by 7.2\% compared to the baselines while incurring an acceptable fidelity penalty ($<10\%$) compared to solo execution.

\subsection{Scheduler}
\label{evaluation:results:scheduler}

{\bf RQ4:} \textit{How well does \projecttitle{}'s scheduler balance fidelity vs. waiting times and balance the load across QPUs?}
We evaluate our scheduler by generating a representative workload consisting of a dataset we collected during the development of \projecttitle{}.

\myparagraph{Dataset collection}
During our exploration of the motivational challenges (\S~\ref{sec:design-challenges}) and experimentation and evaluation of the \projecttitle{} components and their policies, we collected a dataset of $70.000$ benchmark circuits and more than $7000$ job runs in the quantum cloud. We use this dataset to simulate representative workloads, as we detail next.

\myparagraph{Workload generation}
To generate realistic workloads, we monitored all available QPUs on the IBM Quantum Cloud \cite{ibmQuantum} for ten days in November 2023 to estimate the hourly job arrival rate. The average hourly rate is 1500 jobs per hour and is the baseline system workload for our evaluation.

\myparagraph{Fidelity vs. waiting time}
Figure \ref{fig:avg_merged} (a) shows the performance of the formula-based scheduling policy. We show the average fidelity and waiting time as the fidelity weight, $c$, changes (\S~\ref{sec:qos:scheduler}). A weight of $0.7$ achieves $~\sim5\times$ lower waiting times than full priority of fidelity while sacrificing only $\sim2\%$ fidelity.
Figure  \ref{fig:avg_merged} (b) shows the Pareto front of scheduling solutions generated by the genetic algorithm policy. A weight $c=0.5$ achieves $2\times$ lower waiting times with $4\%$ lower fidelity.

\myparagraph{QPU load balancing}
Figure \ref{fig:avg_merged} (c) shows the QPU load as the total runtime each QPU was active, in seconds, for the formula-based policy. All QPUs handle similar loads, with a maximum difference of $15.2\%$.

\myparagraph{RQ4 takeaway}
\projecttitle{} scheduler balances the trade-off between waiting times and fidelity by reducing them 5$\times$ and only 2\%, respectively while balancing the load across QPUs.

\section{Related work}
\label{sec:related}

\myparagraph{Quantum error mitigation}
Error mitigation techniques can be categorized as (1) execution pre- and post-processing \cite{patel2020veritas, das2021jigsaw, patel2020disq, dangwal2023varsaw, tannu2019mitigating, tannu2019ensemble, das2021adapt, patel2021qraft, smoth2022timestitch, patel2022quest, maciejewski2020mitigation}, (2) circuit divide-and-conquer \cite{tang2021cutqc, circuit-knitting-toolbox, Ayanzadeh2023frozenQbits, patel2022quest}, and (3) qubit reuse \cite{hua2023caqr, paler2016wire, bichsel2020silq, paradis2021unqomp, paler2016wire, jiang2024qubit}. Unfortunately, all these techniques are implemented standalone without any infrastructure \coloredText{to compose them}. Instead, \projecttitle{} integrates and composes at least one technique per category in a single software stack, achieving higher fidelity and abstracting away the complexity from the programmer.

\myparagraph{Quantum performance estimation}
Performance estimation in the context of quantum computing exhibits preliminary work, focusing specifically on quantum resource estimation \cite{beverland2022assessingrequirementsscalepractical, quetschlich2024mqtpredictor} and fidelity estimation \cite{mapomatic}. The former work estimates the number of physical qubits required to run an input job given pre-defined fidelity and error correction assumptions. However, \projecttitle{} focuses on the opposite task: given the hardware constraints, it estimates the fidelity of the input quantum job. The latter work has a similar focus and utilizes similar techniques; however, it does not support ML-based approaches or non-analytical-model methods.

\myparagraph{Quantum multi-programming}
Quantum multi-programming work \cite{das2019a, liu2021qucloud} focuses solely on mapping \textit{two} circuits on a single QPU while providing similar fidelity (allocation fairness). Both works overlook the systematic selection of compatible programs for utilization or fidelity and do not account for temporal utilization. Additionally, key optimizations from \cite{das2019a} are integrated into the Qiskit transpiler workflow \cite{qiskit-transpiler}, and therefore, are already used by \projecttitle{}.

\myparagraph{Quantum job scheduling}
Current quantum scheduling methods \cite{ravi2021adaptive, salm2022prioritization, stein2022eqc, wang2024qoncord} are limited because they (1) only perform single-to-many scheduling, (2) do not account for QPU utilization, (3) lack fine control over the fundamental design tradeoffs (\S~\ref{sec:challenges:qpu_load}), or (4) require manual input for final scheduling decisions.
In contrast, \projecttitle{} multiplexes circuits across space and time in a many-to-many fashion, increases QPU utilization, balances the inherent tradeoffs of quantum, and abstracts away the underlying resource management.

\myparagraph{Cloud OSes, resource management, and job scheduling}
Classical cloud operating, resource management, and job scheduling systems have been extensively researched in the past decades \cite{hindman2011mesos, verma2015large, schwarzkopf2013omega, kubernetes, vavilapalli2013apache, yoo2003slurm, sefraoui2012openstack, xiao2018gandiva, peng2018optimus, narayanan2020heterogeneity, kwon2020nimble}. However, the classical domain, even when addressing accelerator heterogeneity, does not face the unique challenges of quantum computing (\S~\ref{sec:design-challenges}) and, as such, cannot be trivially adapted to accommodate its needs.

\section{Conclusion and Discussion}



\coloredText{We presented \projecttitle{}, the first quantum operating system to (1) holistically address quantum computing challenges through a modular, policy-mechanism-separated architecture, (2) enable cross- and intra-layer optimizations via end-to-end co-design, (3) systematically explore key tradeoffs between fidelity, utilization, and waiting times, and (4) introduce novel abstractions and techniques across the stack, including the Qernel, composable error mitigation, compatibility-based multi-programming, and multi-objective scheduling.}

\coloredText{We believe that \projecttitle{} lays the foundation for a quantum operating system whose core principles are designed to persist even as hardware matures. While advances in QPU technology will improve qubit counts and reliability, fundamental challenges—such as noise, heterogeneity, and resource scarcity—will continue to demand careful tradeoffs between fidelity, utilization, and waiting times. Looking forward, \projecttitle{}'s principles extend naturally to fault-tolerant quantum computing (FTQC): \projecttitle{}’s fidelity-aware scheduler and compatibility-driven multi-programmer can be adapted to logical qubit placement, routing, and error-corrected resource allocation, preserving execution efficiency while accounting for FTQC-specific overheads.}

\section*{Acknowledgements}

We sincerely thank our shepherd and the anonymous reviewers for their feedback. We thank Karl Jansen and Stefan Kühn from the Center for Quantum Technology and Applications (CQTA)- Zeuthen for supporting this work by providing access to IBM quantum resources. We also thank Ahmed Darwish and Dmitry Lugovoy for their contributions to this work. Funded by the Bavarian State Ministry of Science and the Arts as part of the Munich Quantum Valley (MQV).

\typeout{}
\bibliographystyle{plain}
\bibliography{references}

\end{document}